\documentclass[%
 reprint,
 amsmath,amssymb,
 aps,
]{revtex4-1}
\usepackage{xcolor}
\newcommand{\ie}{\textit{i.e.}}
\usepackage{graphicx}
\usepackage{dcolumn}
\usepackage{bm}

\begin{document}


\title{Public disorder and transport networks in the Latin American Context}

\author{Carlos Cartes}
 \email{corresponding author: carlos.cartes@miuandes.cl}
\affiliation{%
 Universidad de los Andes, Chile. \\ Complex Systems Group, Facultad de Ingenier\'{\i}a y Ciencias Aplicadas.
}%


\author{Toby P. Davies}
\affiliation{
 Department of Security and Crime Science, University College London, \\
 35 Tavistock Square, London WC1H 9EZ, UK
}%

\date{\today}

\begin{abstract}

We propose an extension of the Davies et al. model, used to describe the London riots of 2011. This addition allows us to consider long travel distances in a city for potential rioting population. This is achieved by introducing public transport networks, which modifies the perceived travel distance between the population and likely targets. Using this more general formulation, we applied the model to the typical Griffin and Ford pattern for population distribution to describe the general features of most large Latin American cities. The possibility of long-range traveling by part of the general population has, for an immediate consequence, the existence of isolated spots more prone to suffer from rioting activity, as they are easier to reach than the rest of the city. These areas finally made it easier to control the eventual disorder by part of police forces. The reason for this outcome is that transport networks turn riots into highly localized and intense events. They are attracting a large police contingent, which will later extinguish the remaining disorder activity on the rest of the city. Therefore, working transport networks in a city effectively reduces the number of police force contingent required to control public disorder.
This result, we must remark, is valid only if the model requisites for order forces are satisfied: extra police contingent can be added swiftly as required, and these forces can move around the city with total freedom.

\begin{description}
\item[PACS numbers]
\end{description}
\end{abstract}

\pacs{Valid PACS appear here}
\maketitle


\section{\label{sec:level1}Introduction}

Public disorder, riots and civil unrest are all manifestations of collective behaviour in which hostile action is directed towards members belonging to other groups or authorities. If mishandled, such phenomena can sometimes cascade into large--scale outbreaks of violent criminality \cite{salehyan2012social}. This is a significant public security problem, which has substantial consequences in terms of both personal injuries and property damage. Furthermore, if disorder escalates into a nation-wide episode, it can pave the way for political instability or even the overthrow of governments, as in Tunisia and Egypt during the Arab Spring of 2011.

The ability to understand and predict the spatial and temporal evolution of disorder is, therefore, of huge importance for policy. As with other forms of crime, such understanding forms a basis for preventative interventions, and allows resources (such as policing) to be used in a way that mitigates negative consequences most efficiently. This need is particularly acute for rare events such as riots, since the lack of real-world examples means that authorities have little opportunity to develop their understanding or evaluate approaches empirically. For this reason, mathematical modelling and numerical simulation has great potential as a means of developing insight. Not only can they offer profound understanding of the phenomena themselves, but they can be used as a means of exploring alternative scenarios; that is, performing `what if' experiments for riot situations. In particular, this provides a means of testing potential control or mitigation strategies at low cost and at a high level of granularity.

Early analyses considered collective disorder to be a return to a ``natural condition'', whereby crowds temporarily forgot civilized behavior and instead regressed to more primitive ways. This phenomenon was referred to as the ``madness of the crowds'' \cite{le2017crowd} and ``mass enthusiasm'' \cite{lorenz2005aggression}. The alternative suggestion that conscious and rational decisions were involved in the participation in riots and public disorder began with the pioneering work of Granovetter \cite{granovetter1978threshold, Granovetter1983}. In the models presented, potential rioters evaluate the benefits and costs of joining public disorder, taking into account the social influence of other individuals. Later developments, such as Epstein's formulation for social conflict \cite{epstein2002modeling} using Agent--Based Models (ABM), successfully described more detailed classical patterns in public disorder, such as the long calm periods interrupted by intermittent bursts of collective violence. Further approaches to this problem were made by using partial differential equations, capturing important features like the contagious nature of rioting behavior and reflecting the observed patterns from the French riots of 2005 \cite{berestycki2015model, bonnasse2018epidemiological}. Also using differential equations, Davies et al. \cite{Davies2013} successfully captured the quantitative dynamics of the London riots of 2011. This model was based on empirical evidence from London that rioters chose target locations on the basis of their characteristics \cite{baudains2013target}; such behaviour is consistent with the `rational choice' perspective on criminal decision-making \cite{cornish_reasoning_1986}, in which offenders are assumed to select potential targets from a range of alternatives in such a way that they maximise reward and minimise risk \cite{bernasco2004residential, ruiter_crime_2017}. In the Davies et al. model, the participation of rioters was influenced by the volume of retail activity (\ie\@ potential for looting) in possible target locations, as well as the number of other participants already present (as a measure of `safety in numbers').

Although spatial data has been incorporated in riot models, it has typically been simplistic and included little information concerning urban structure. In particular, there has been no attempt to consider the effect of transport networks, in the form of either public transport or simply streets themselves. This constitutes a significant shortcoming: transport networks have been shown to have a significant effect in promoting crime in general \cite{brantingham1991public, swartz2000spatial, davies_examining_2014} by either generating new opportunities for offending or increasing the accessibility of existing ones. Indeed, analysis of the London riots indicates a strong correlation between the probability of a location being looted and the proximity of underground train stations \cite{baudains2016london}. In this work, we seek to extend the previous model of Davies et al by incorporating the effect of a transport network, in order to increase the validity and granularity of the model. Inspired by the recent outbreaks of violence in Santiago, Chile, and Latin American cities from Bolivia, Peru, Colombia, Equador among others \cite{TheConversation2019, bjork2020mass, WashingtonPost2020, TheGuardian2019, TheGuardian2019-2},  we do this with particular focus on the context of Latin American cities.

We propose to implement the typical layout of a Latin American city, as a prototype transport network and geographic configuration, to test our formulation. A stylised model of such cities was first proposed by Griffin and Ford in 1980 \cite{griffin1980model}. Briefly, it formulates cities as being structured around a ``spine'', where wealthy inhabitants live, surrounded by three concentric regions in which infrastructure, public services and quality of life in general decreases with distance to the city center. This model was later revised by Ford in 1996 \cite{ford1996new}, adding some novel elements but maintaining its principal characteristics. Although the steady decline in Latin American population growth since the 1980's has led to a constant improvement in the quality of public services and amenities for the inhabitants of more distant zones, showing evidence that these cities are converging towards an Anglo--American configuration \cite{borsdorf2003como}, Buzai stated recently \cite{buzai2016urban} that Ford's 1996 model remains valid.

Of relevance for the case of the public disorder occurring in Latin American cities, a number of recent articles have presented quantitative analysis of the Chilean riots of 2019. For instance, Caroca et al. \cite{Caroca2020} characterized the temporal evolution of the total intensity of the events, showing that it follows the typical curve predicted by Burbeck \cite{burbeck1978dynamics} (rapid escalation to a pronounced peak, followed by exponential decay). The spatial distribution of activity on the first days for the capital city Santiago was also characterized by Cartes et. al. \cite{Cartestransport}, with the analysis showing that activity appeared to be concentrated primarily around subway stations. Nevertheless, we are not aware of any research concerning policies for control and mitigation applied to those particular scenarios.

In this paper, we present the Davies et al. model used to describe the London riots of 2011 and extend it to incorporate the effects of public transport networks on the underlying spatial environment. These modifications mean that riot participants are able to travel large distances (and therefore reach distant targets) at low cost, in a way that they would not otherwise be able to do. We then implement the model with the typical population distribution of a Latin American city. Depending on the accessibility of the public transport network and the police response’s effectiveness, we identify different possible scenarios regarding the intensity of the riots and their distribution in the city.

This document is organized in the following way. In \ref{sec:level2} we introduce Davies et al. model for riots, its extension to incorporate public transport networks, and the considerations taken on the context of Latin American cities. In \ref{sec:results} we present in detail the model’s results, specifically the intensity of the riots and its spatial distribution. At \ref{sec:Analysis} we analyze the results presenting our conclusions about the model, and finally, in \ref{sec:Perspectives} we show our perspectives and future work.

\section{\label{sec:level2} The Model}

\subsection{Davies et al. (2013) formulation}

The model introduced by Davies et al. is situated on a discrete spatial system comprising two sets of locations: residential areas, indexed by $i$, and retail sites (the targets for rioting), indexed by $j$. In essence, the model seeks to describe the flows of rioters from their homes to retail sites, and their interactions with police once there. The model tracks the numbers of individuals in each location over time, with the key quantities being $R_j(t)$ and $P_j(t)$: respectively the numbers of rioters and police in location $j$ at time $t$. In mechanistic terms, the model has three stages, corresponding to different phases of an individual's involvement in rioting:
\begin{itemize}
    \item The decision to participate in rioting
    \item The choice of site at which to offend
    \item Possible arrest due to interaction with police
\end{itemize}
In the model, citizens pass through three possible states - \textit{inactive}, \textit{participating} and \textit{arrested} - so that the overall model is analogous to the Susceptible-Infected-Removed epidemiological approach.

For the first two of these stages, a single quantity - referred to as \textit{attractiveness} - plays a central role. The attractiveness of a retail site $j$, from the perspective of a citizen in residential area $i$, denoted $W_{ij}$, represents the result of a cost/benefit calculation for individuals considering the utility of rioting in location $j$. It takes into account three factors: the volume of retail activity at $j$ (benefit), the distance between $i$ and $j$ (cost) and the probability of arrest when rioting at $j$ (cost/risk). It is calculated as:


\begin{equation}
  \label{eq:benefit}
  W_{ij}(t) = Z_j^{\alpha_r}\exp\left( -\beta_r d_{ij} \right) \exp\left( - \left\lfloor \frac{\gamma_r P_j(t)}{R_j(t)}\right\rfloor \right)
\end{equation}
\noindent where $Z_j$ is the volume of retail activity at $j$, $d_{ij}$ is the Euclidean distance between $i$ and $j$ (the minus sign reflects that it acts as an impedance) and the third term $\left\lfloor\frac{\gamma_r P_j(t)}{R_j(t)}\right\rfloor$ corresponds to the probability of arrest at $j$ (a ``deterrence''). This is proportional to the ratio between police officers $P_j(t)$ and rioters $R_j(t)$, with the floor function $\lfloor \rfloor$ added to account for the empirical fact that when police are outnumbered ($\gamma_r P_j(t) < R_j(t)$) the disorder becomes out of control and police are incapable of making arrests \cite{wilensky2004netlogo}. Finally the constants $\alpha_r$, $\beta_r$ and $\gamma_r$ are free parameters to calibrate the model by using the available data.

The first step in the model concerns the decision of whether to participate in rioting. This is determined by a `probability of offending', defined for each residential area, $i$, and representing the probability that residents of that area will choose to participate. This is modelled as a sigmoidal function, based on the total attractiveness of all possible targets $j$, as perceived by residents of $i$:

\begin{equation}
  \label{eq:prob_off}
  P_i^{\rm offend} = \rho_i^{\mu}\frac{\sum_j W_{ij}(t)}{ 1 + \displaystyle{\sum_j W_{ij}(t)}} 
\end{equation}

\noindent The multiplicative factor $\rho_i$ is a measure of the deprivation at $i$, reflecting the empirical finding that higher levels of deprivation are associated with greater participation \cite{Davies2013, baudains2013target}. Taking the product of this probability with the number of inactive residents of $i$, denoted $I_i(t)$, gives the number of new crowd participants:

\begin{equation}
  \label{eq:actvN}
  N_i(t) = \eta I_i(t) P_i^{\rm offend}(t)
\end{equation}

\noindent where $\eta$ is a transition rate.

Once a citizen decides to participate, the next stage of the model concerns the choice of location. This is modelled via the use of an entropy--maximising spatial interaction model (SIM), which estimates the most probable flows between a set of origins and destinations, given known out-flows and an expression for the utility of each destination \cite{wilson1971family}. The utility used is the \textit{attractiveness} defined in \eqref{eq:benefit} above; however, it is first modified by taking a moving average, in order to account for the delay in reaction of participants. This represents both travel time between origins and destinations and the time taken for information about the riot situation to propagate through the population. This temporal average of $W_{ij}(t)$ over the most recent $L_r$ time--steps gives the ``effective attractiveness'' between $i$ and $j$:

\begin{equation}
  \label{eq:attract}
  W_{ij}^e = \frac{1}{L_r}\sum_{l = 0}^{L_r - 1}W_{ij}(t - l\delta t)\,.
\end{equation}

Now if $A_i(t)$ denotes the number of residents of $i$ who are active at time $t$, then the number of those participating in rioting at $j$ is given by \cite{wilson2007boltzmann}

\begin{equation}
  \label{eq:rioterflow}
  T_{ij}(t) = \frac{A_i(t) W_{ij}^e}{\displaystyle{\sum_k W_{ik}^e}},
\end{equation}

\noindent so that the rioters from $i$ are assigned to targets in proportion to their attractiveness. By summing the above expression over all origin locations $i$, the total number of rioters located at the site $j$ can be obtained:

\begin{equation}
  \label{eq:rioters}
  R_j (t) = \sum_{i} \frac{A_i(t) W_{ij}^e}{\displaystyle{\sum_j W_{ij}^e}}\,.
\end{equation}

Simultaneously with the flows of rioters, the locations of police forces are also modelled. These are done in a similar way to the modelling of riot behaviour, though with two differences: the number of police officers remains constant, and an alternative utility function - not including distance as an impedance factor - is used. This utility - referred to as  \textit{demand} - depends on the value of property at a site (again modelled as retail volume, $Z_j$) and the number of rioters $R_j$:

\begin{equation}
  \label{eq:requirementorig}
  D_j = Z_j^{\alpha_p} \exp \left(\gamma_p R_j(t) \right)
\end{equation}

\noindent Here $\alpha_p$ and $\gamma_p$ are parameters used to calibrate the model. In the same way as the attractiveness, the ``effective demand'' is computed by averaging over the most recent $L_p$ time--steps

\begin{equation}
  \label{eq:effrequirement}
  D_{j}^e(t) = \frac{1}{L_r} \sum_{l = 0}^{L_p - 1}D_j(t - l \delta t)\,,
\end{equation}

\noindent again reflecting a reaction delay. Police officers are then assigned to each location $j$ according to the proportion of total effective demand; with $P^{\rm total}$ as the total number of officers, $P_j(t)$ is given by:

\begin{equation}
  \label{eq:poldist}
  P_j(t) = P^{\rm total}\frac{D_{j}^e(t)}{\displaystyle{\sum_k D_{k}^e(t)}}\,.
\end{equation}

The final stage of the model concerns the arrest of rioters by police. For each residential location $i$, the number of residents arrested in $j$ is obtained by multiplying the flow $T_{ij}(t)$ by the arrest rate in $j$. Taking the sum over all riot sites gives the total number of arrests of residents of $i$:

\begin{equation}
  \label{eq:capture}
  C_i(t) = \tau \sum_j T_{ij}(t)\left(1 - \exp\left( -\left\lfloor\frac{P_j(t)}{R_j(t)}\right\rfloor\right) \right)\,.
\end{equation}

The above processes are combined to give a dynamical system tracking the active and inactive populations of each residential site $i$:

\begin{eqnarray}
  \label{eq:dynamicsystem}
  A_i(t  + \delta t) & = & A_i(t) + \delta t \left(N_i(t) - C_i(t) \right) \\
  I_i(t  + \delta t) & = & I_i(t) - \delta t N_i (t) \,. \nonumber
\end{eqnarray}

\subsubsection{Addition of transport networks}
In this paper, we introduce a modification to the above model, incorporating the effect of public transport networks. Before doing this, we first simplify the spatial domain on which it is defined: rather than an arbitrary spatial system, we assume that the model is defined on a regular two-dimensional grid. This can be interpreted as a `rasterised' version of a generic urban area, and therefore represents a generalisation of the model. In this formulation, we assume that all grid cells represent residential areas, while retail sites (\ie\@ targets) are located at a subset of grid points. The model defined above translates straightforwardly to this setting, with the distance between any two locations defined as the Euclidean distance between the grid points.

In incorporating transport networks, the central principle is that the effect of a transport line is to reduce the travel cost along the paths it serves. We model a network as a set of nodes, representing stations or stops, with connections between them representing direct transport links. The effect of the transport line is modelled by applying a multiplicative factor $\delta < 1$ to the distance between any pair of connected nodes: the link therefore acts as a `shortcut' relative to regular travel. For any two neighbouring nodes, the distance is therefore given by 

\begin{equation}
  \label{eq:dist_delta}
  d = \delta\sqrt{\Delta x^2 + \Delta y^2}\,.
\end{equation}

For any pair of nodes $v$ and $w$ on the network, the travel distance $d^n_{vw}$ is given by the sum of these distances over all constituent network links; essentially the Euclidean length of the path, multiplied by $\delta$.

Once the transport network has been implemented, the distances between all origins and destinations are updated to take its effect into account. In particular, for any origin $i$ and destination $j$, the distance when using the transport network is computed on the following basis:

\begin{enumerate}
    \item The distance $l_1$ from $i$ to the nearest transport node $v$ is calculated
    \item The distance $l_2$ from $j$ to the nearest transport node $w$ is calculated
    \item The total effective distance is given by $d_{ij} = d^n_{vw} +l_1 + l_2$
\end{enumerate}

For the purpose of the model, the distance $d_{ij}$ is then defined as the minimum of the Euclidean distance and transport distance.

\subsection{Latin American Cities}

\begin{figure}[h!]
\begin{center}
\includegraphics[width=8.cm,angle=0]{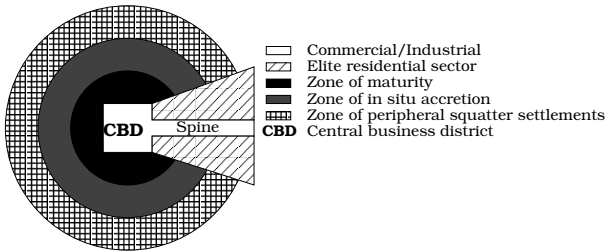}
\caption{Model for a Latin American City, after Ford and Griffin 1980.} 
\label{fig:lat_am_city}
\end{center}
\end{figure}

We propose to implement the typical layout of a Latin American city, as a prototype transport network and geographic distribution pattern to test our formulation. A generic model of such cities was proposed by Griffin and Ford in 1980 \cite{griffin1980model}, consisting of a ``spine'' where wealthy inhabitants live, surrounded by three concentric regions, over which infrastructure, public services and quality of life in general decrease with distance to the city center. The names of these regions are, from center to periphery: zone of maturity, zone of in situ accretion and zone of peripheral squatter settlements; see FIG. \ref{fig:lat_am_city}. This configuration is an evolution from the colonial city \cite{gilbert1987urban} and is predominantly shaped by the migration to cities from rural areas that occurred during the 1940s and 1950s. The fast increase in urban population, under which new migrants located themselves at the borders of the city, generated a variety of problems as housing, public services and employment opportunities could not keep pace with their need. Faced with these constraints, city planners expanded public services like electricity, sewer systems and paved streets in only one direction, with this preferred orientation forming the `spine'. Naturally, all important urban facilities like banks, parks and restaurants - along with high--status residential housing - concentrated around that region, in order to make use of all the aforementioned public services. Population density within the various regions of the city differs dramatically, with upper class groups having the lowest density, middle class occupying the areas with higher density and lower income groups having densities in the middle of the two \cite{amato1970comparison}. This model was later revised by Ford in 1996 \cite{ford1996new}, who added some novel elements while maintaining the principal characteristics. The steady decline in population growth in Latin America since the 1980s has alleviated some of the issues facing major cities, with constant improvement on the quality of public services and amenities for the inhabitants of the further away zones. Nevertheless, although there is some evidence that these cities are converging toward an Anglo--American configuration \cite{borsdorf2003como}, Ford's 1996 model remains valid in many contexts \cite{buzai2016urban}.

It is not surprising that this population distribution inside the city tends to produce high levels of segregation and social friction among the citizens, since all groups are constrained to occupy a strictly bounded area, sharply delimited by income and social status. Evidence of urban segregation is found in Mexico \cite{schteingart2001division}, Argentina \cite{torres2001cambios}, Brazil \cite{villacca1998espacco} and Chile \cite{sabatini2001segregacion}. Mobility studies from the latter country's capital city, Santiago, found that even movements inside the city are restricted to individuals' designated residence areas \cite{dannemann2018time}, where they remain isolated, with the only common meeting point in the whole urban area being the center of the city.

\subsection{Initial population and target distribution }

For our population arrangement, we used the income distribution indicated in Ford's model. On the other hand, our model’s population density comes from the values mentioned above shown by Amato \cite{amato1970comparison}. All our simulations were carried out with a total population of $1.2$ million inhabitants.

\begin{figure}[h!]
\begin{center}
\includegraphics[width=8.cm,angle=0]{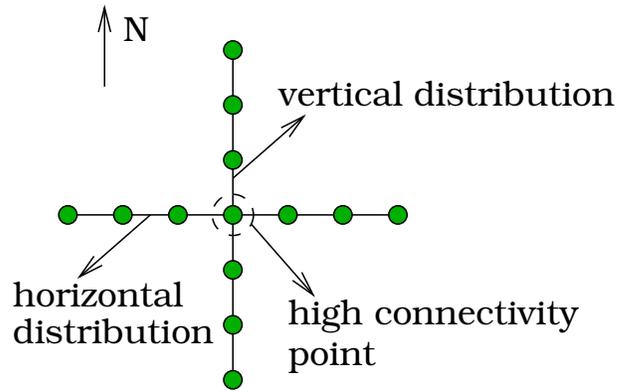}
\caption{Scheme for the transport network implemented in the simulations. It bridges the suburbs with the CBD, creating
a high connectivity point. Network nodes and targets share spatial location (green dots).} 
\label{fig:Transport_netw}
\end{center}
\end{figure}

Our simplified transport network consists of only two lines intersecting at the center of our environment, coincidentally with the city’s Central Business District; see FIG. \ref{fig:Transport_netw}. This configuration is the simplest one that allows high mobility for the inhabitants of the periphery, creating a direct link with the city center. This arrangement also creates a focal location with relatively high connectivity, easily reached from all the city’s regions.

\begin{figure}[h!]
\begin{center}
\includegraphics[width=8.cm,angle=0]{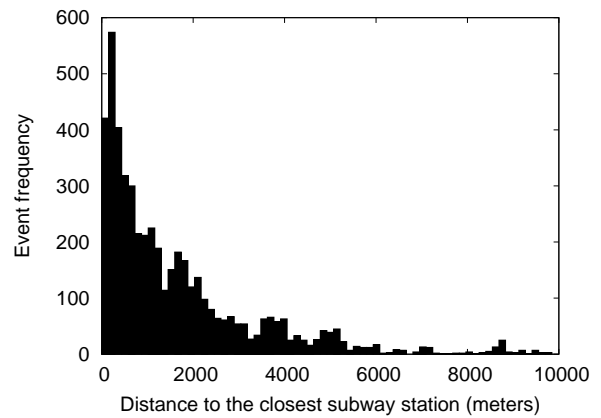}
\caption{Event frequency as a function of the distance to subway stations is shown. Around half of the total lies within $1$km or less from a subway station.} 
\label{fig:Event_Distribution}
\end{center}
\end{figure}

The spatial distribution of targets for disorder activity was intentionally made by sharing location with the transport network nodes, see FIG. \ref{fig:Transport_netw}. This choice was taken because highly connected spots, like subway stations, attract retail activity \cite{Bowes2001, debrezion2004impact}. It was also found that during the first four days of the Chilean riots from October 2019, in the capital city Santiago, most of the events were concentrated around subway stations (called Metro Network) \cite{Cartestransport}. In FIG. \ref{fig:Event_Distribution} the frequency of events, as a function of the distance to the closest subway station, is shown, with the distribution following approximately a power law. It is evident that most of the disorder activity was concentrated around these transport nodes: for example, around $50\%$ of the total activity took place $1$km or less from these well-connected spots \cite{Cartestransport}. We also take into consideration that there is a correlation between commercial and residential real estate values \cite{schoenmaker2015real}. Therefore, the perceived attractiveness of these spots increases towards the center of the high-income region. On the other hand, the targets decrease in value as they are located farther away from the center, entering the peripheral low-income regions.

Indeed it is possible to argue that, at least initially, disorders aren't motivated by a desire to make a monetary profit from the looting opportunity; instead, they are triggered on political and social grounds. Nevertheless, Santiago’s most affected area is effectively the most connected spot on the whole city \cite{Cartestransport}, the reasons for which are likely to be its easier access and the awareness of its existence by the population. We consider it essential to remark that this was not the first time this particular city spot suffered large-scale disorders. Historically that region was the default meeting point for any kind of massive concentration, independent of the motivation, being it the positive result in a sport like Copa Am\'erica in 2015 \cite{Emol2015} or a political one, like the student demands for improvements on education policies in 2011 \cite{Ruz2019}.

\subsection{Numerical implementation}

For our simulation, we used a grid comprised of $64\times 64$ elements to contain the population, and another grid, with the same dimensions, was used to contain the targets. According to FIG \ref{fig:lat_am_city}, the initial population is completely inactive, and its density at any location depends on the sector in which it lies. The same distribution applies to deprivation. The actual values are shown in TABLE \ref{table:population}.

\begin{table}[h!]
\begin{center}
 \begin{tabular}{||c | c | c||} 
 \hline
 Sector & Pop.$/$Element & Deprivation \\ [0.5ex] 
 \hline\hline
 Elite residential & 240 & 0.01 \\ [0.5ex]
 \hline
 Zone of Maturity & 520 & 0.05 \\[0.5ex]
 \hline
 Zone of in situ & 400 & 0.1 \\
accretion & & \\[0.5ex]
 \hline
 Zone of peripheral & 328 & 1 \\
squatter settlements & &\\[0.5ex]
 \hline
\end{tabular}
\end{center}
\caption{Population and deprivation values per grid element, used on the simulations, each sector corresponds to those shown in FIG. \ref{fig:lat_am_city}}
\label{table:population}
\end{table}

The numerical parameters in our simulations are the same used by Davies et al. \cite{Davies2013} as they proved to be helpful to describe an actual rioting situation. The only exception is $dt$ and the temporal delays $L_r$ and $L_p$, as they were increased to slow down the reaction time for both rioters and police forces. This choice was adopted to take into account the larger distances involved in our simulations. The actual values implemented are shown in TABLE \ref{table:parameters}. We must remark that temporal dynamics is preserved as long as the products $L_r dt$ and $L_p dt$ are kept constant. In this way, we can increase $dt$ and study the long-term evolution of the system, or lower it to examine the initial dynamics in more detail.

\begin{table}[h!]
\begin{center}
 \begin{tabular}{c | c | c| c | c } 
 
 $\alpha_r = 0.6$ & $\beta_r = 0.5$ & $\gamma_r = 0.11$ &  $L_r = 28$ & $dt = 0.25$ \\ [0.5ex]
 \hline\hline
 $\alpha_p = 0.65$ & $\gamma_p = 0.012$ & $L_p = 40$ & $\eta = 0.006$  & $\tau = 0.75$
 
\end{tabular}
\end{center}
\caption{Numerical values of the parameters used for all the simulations.}
\label{table:parameters}
\end{table}

\subsection{Police forces behavior}

To explain how activities of police forces are implemented, first we must recall one of the premises of the model: because of eq. (\ref{eq:capture}), every combination on the number of rioters and police forces, at any given region in space, can conduce to only two different and mutually excluding scenarios: i) Police forces regain complete control of the situation and the incident is immediately extinguished, or ii)  Police forces are overwhelmed and cannot take back control of the situation, without the addition of reinforcements. This dichotomous behavior depends on the initial number of officers and is a byproduct of the floor function present at the capture rate. When $P_j(t)<R_j(t)$, police forces completely lose control of the situation, since the floor function forbids to perform any detention, by making the capture probability nil.

 Because riot formation requires a lesser number of police forces than rioters, at any given location, we chose that $P_{total} = \alpha_1 R_{total}$ . Where $R_{total} = \displaystyle{\sum_j R_j}$ is the total number of rioters and $\alpha_1 < 1$  is a constant, smaller than one, if the total number of rioters remains less than a threshold value, this means $R_{total} < R_C$, thus promoting conditions for riot formation. Once this critical value is reached, the total number of police forces becomes $P_{total} = \beta_1 R_{total}$, where  $\beta_1 > 1$, allowing the possibility to regain global control of the situation by police forces. As an additional condition, we also imposed that when the maximum value for $P_{total}$ is reached, it is kept constant, making the total rioting activity on the city a lot easier and faster to control. If this last condition is not kept, control is lost for a period almost $10$ times larger than otherwise.

\section{\label{sec:results}Results}

\subsection{Rioter Distribution}

The new parameters introduced in our current formulation can potentially change the system dynamics. To explore this, we examined different combinations of the constant values in the model set-up. By changing those parameters, we found that the result can develop in different ways by changing:

\begin{itemize}
    \item $\delta$: Accessibility, lower values mean equally lower travel times, $1$ represents walking to a destination.
    \item $R_C$: Riot critical intensity.
    \item $\beta_1$: Multiplicative factor of order forces, after the critical intensity $R_C$ is reached.
\end{itemize}

All other parameters were kept constant: for example, the value $\alpha_1 = 0.7$ was maintained because different values did not change significantly the dynamics of the simulations, only slightly affecting the time at which $R_C$ is reached (and not modifying the posterior riot evolution).

\begin{figure}[h!]
\begin{center}
\includegraphics[width=8.cm,angle=0]{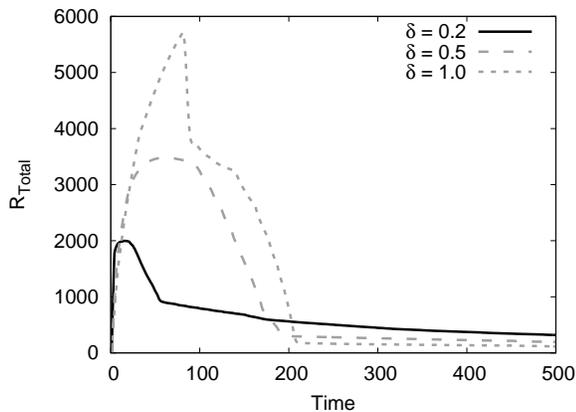}
\caption{Evolution of the total number of rioters $R_{\rm Total}$, as function of time. For the parameters $\beta_1 = 1.5$, $R_C=1500$ and three different values of $\delta$.} 
\label{fig:rioters_pl_1.5_RC_1500}
\end{center}
\end{figure}

\begin{figure}[h!]
\begin{center}
\includegraphics[width=8.cm,angle=0]{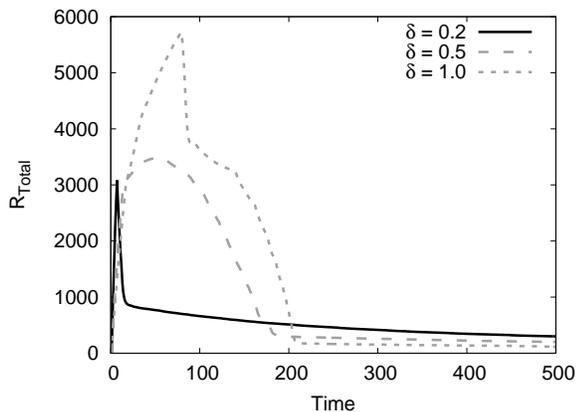}
\caption{Evolution of the total number of rioters $R_{\rm Total}$, as function of time. For the parameters $\beta_1 = 1.5$, $R_C=3000$ and three different values of $\delta$.} 
\label{fig:rioters_pl_1.5_RC_3000}
\end{center}
\end{figure}

Further exploration of the simulation parameter space, for example $R_C$ and $\beta_1$, indicates that changes performed on the former make an appreciable difference only when $\delta$ is small, corresponding to a situation in which it is easy for the active population to move around the city; see FIGS. \ref{fig:rioters_pl_1.5_RC_1500} and \ref{fig:rioters_pl_1.5_RC_3000} with $\delta = 0.2$. In the case when $R_C = 1500$, rioter accumulation has a certain ``inertia'', due to the relative small police force contingent and the proportionately high active population ready to join the disorder. In that situation the peak value of $R_{total}$ is around $2000$, see FIG. \ref{fig:rioters_pl_1.5_RC_1500}. On the other hand, when the threshold value is increased to $R_C = 3000$, most of the activity is immediately extinguished once that value is reached, leaving only a scattered remaining as shown in FIG. \ref{fig:rioters_pl_1.5_RC_3000}.

The floor function in eq.  (\ref{eq:capture}) controls the influence of the second parameter, $\beta_1$, on the dynamics. Different values of this parameter can therefore only bring about different results when the increments are in the order of $1$. In consequence, when $\beta_1$ lies within on the range $1<\beta_1<2$ all simulation results are similar, in view of the fact that the capture rate is more or less the same. If this control parameter is further increased, with $\beta_1>2$, the outcome for any value of $R_C$ and $\delta$ is very close to the result shown in FIG. \ref{fig:rioters_pl_1.5_RC_3000} with $\delta = 0.2$. This behavior implies a quick extinction of the intense rioting activity, leaving a scattered residual disorder. Therefore, in a riot situation, control can be regained by order forces when roughly $P_{total} \sim 2 R_{total}$. It is a strong indication that control can be taken back very quickly if all available order forces are deployed as fast as possible once the threshold $R_C$ is reached. While the ratio $P_{total}/R_{total}$ is the largest possible, riots will have a shorter duration and reduced intensity.

\subsubsection{Low accessibility results $\delta=1$}

Here we will describe the spatial distribution of disorder activity when there is low accessibility. In this case, there is no available public transport network, and therefore any movements by the active population or potential rioters are made by walking. In consequence, any rioting activity will tend to be within a close distance of the participant’s residence.  

\begin{figure}[ht]
\begin{center}
\includegraphics[width=8.cm,angle=0]{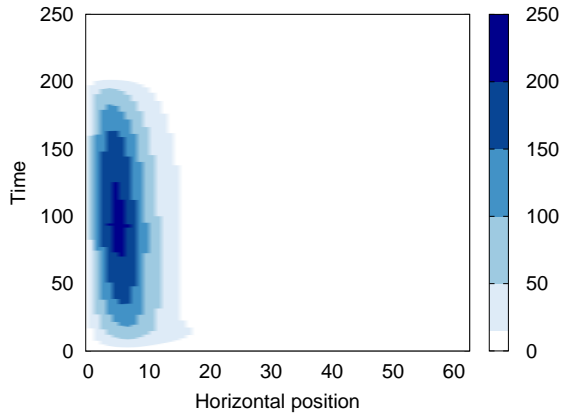}
\caption{Horizontal rioter distribution for $\beta_1 = 1.5$, $R_C=3000$ and $\delta = 1.0$.} 
\label{fig:rioters_pl_1.5_RC_3000_delta_1.0_horizontal}
\end{center}
\end{figure}

\begin{figure}[ht]
\begin{center}
\includegraphics[width=8.cm,angle=0]{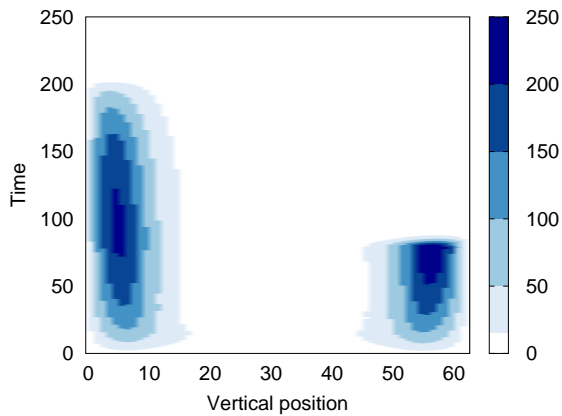}
\caption{Vertical rioter distribution for $\beta_1 = 1.5$, $R_C=3000$ and $\delta = 1.0$.} 
\label{fig:rioters_pl_1.5_RC_3000_delta_1.0_vertical}
\end{center}
\end{figure}

Because of these mobility restrictions, rioting activity is generated and remains confined to the city’s peripheral zones. It seems that the origin of the disorder is located at the boundary between the two outer regions, which are also the most deprived areas of our configuration. The fact that a division between two regions increases the possibility of conflict is in agreement with \cite{rutherford2014good}, where these particular spots, localized at the boundaries, generally have a historical record of public disorder and civil violence,  as shown in FIGS. \ref{fig:rioters_pl_1.5_RC_3000_delta_1.0_horizontal} and \ref{fig:rioters_pl_1.5_RC_3000_delta_1.0_vertical}.

On the opposite side of the city, high-income areas are virtually conflict-free, as should be expected, because of their low levels of deprivation. The scarce activity is concentrated at the center of our configuration, where the frontier with other regions occurs.

On the other hand, at the peripheral regions, riots are widespread, summoning large fractions of the population overall the surface surrounding the target region. We also found a high asymmetry on the vertical distribution of rioters, shown in FIG. \ref{fig:rioters_pl_1.5_RC_3000_delta_1.0_vertical}; in fact, the southern half of the city has a riot activity that lasts almost two times more than the northern counterpart. The difference is a byproduct of a slight asymmetry in the population and deprivation initial distributions, being the north half marginally more significant than the south. The relative difference lies in the order of $1/32$.

The result indicates that small and almost negligible differences in the initial conditions have as in consequence large deviations in the disorders’ long-term behavior. In this particular case the total duration of the activity in the southern region has half the duration of that within the northern region. Furthermore, it is more intense, recruiting a larger number of rioters in less time. This localized increase in the number of rioters will attract more police forces, leading to a faster extinction of the activity once the threshold $R_C$ is reached. This excess of forces will be displaced subsequently to the remaining areas with active conflicts. This movement can be very slow, only depending on the equations’ delay term, as shown in eq. (\ref{eq:effrequirement}).

\subsubsection{High accessibility results $\delta=0.2$}

\begin{figure}[ht]
\begin{center}
\includegraphics[width=8.cm,angle=0]{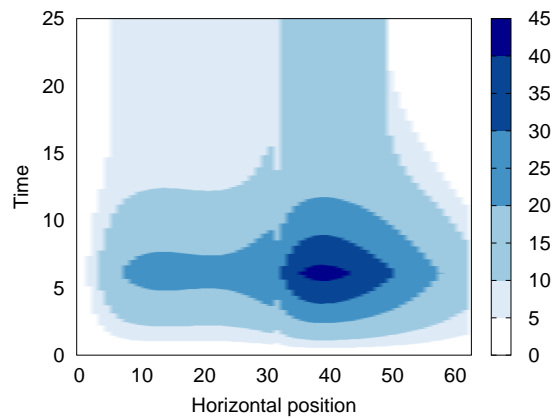}
\caption{Horizontal rioter distribution for $\beta_1 = 1.5$, $R_C=3000$ and $\delta = 0.2$.} 
\label{fig:rioters_pl_1.5_RC_3000_delta_0.2_horizontal}
\end{center}
\end{figure}

\begin{figure}[ht]
\begin{center}
\includegraphics[width=8.cm,angle=0]{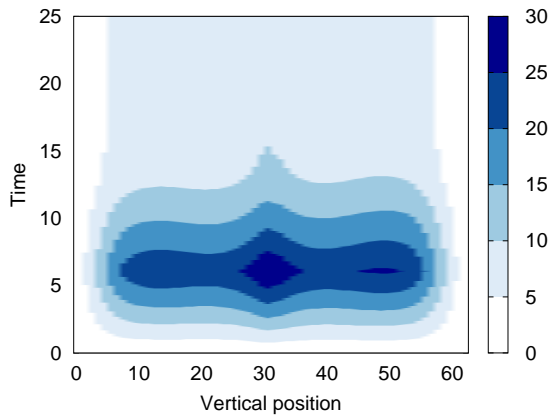}
\caption{Vertical rioter distribution for $\beta_1 = 1.5$, $R_C=3000$ and $\delta = 1.0$.} 
\label{fig:rioters_pl_1.5_RC_3000_delta_0.2_vertical}
\end{center}
\end{figure}

In this case, the origin of rioter activity is the intersection point of the transport network. From there, the disorder expands towards the peripheral regions of the city. This expansion has the shape of travelling waves, similar to those found by Berestycki \cite{berestycki2015model, bonnasse2018epidemiological, berestycki2020modeling}; unlike in that case, however, these waves are not characterized by a constant speed, but instead the ``disorder front'' slows down as the riot reaches the periphery of the city, see FIGS. \ref{fig:rioters_pl_1.5_RC_3000_delta_0.2_horizontal} and \ref{fig:rioters_pl_1.5_RC_3000_delta_0.2_vertical}. The extinction of rioter activity, a consequence of police action, also shows a traveling wave pattern with a lower speed.

It is important to remark that the peak of rioter activity of the horizontal distribution is not exactly located at the intersection. Still, it is shifted towards the elite residential sector. This phenomenon is explained by the balance between more attractive targets for the population and the traveling distance. Another high activity spot is located in the opposite direction but still close to the Central Business District. Nevertheless, the larger portion of rioting activity lasts, comparatively, for a very short time. There are still some residual rioter groups that stay active for a longer time, but they are scattered widely across the domain in a way that is homogeneous over the vertical part of the domain and highly asymmetric on the horizontal segment; see FIGS. \ref{fig:rioters_pl_1.5_RC_3000_delta_0.2_horizontal} and \ref{fig:rioters_pl_1.5_RC_3000_delta_0.2_vertical}. Therefore in this case some activity is preserved, on the long term, at the most deprived regions of the city. Nevertheless, peak rioting activity is at least five times lower. Its duration is eight times shorter than the case without a transport network, comparing  FIGS. \ref{fig:rioters_pl_1.5_RC_3000_delta_1.0_horizontal} and \ref{fig:rioters_pl_1.5_RC_3000_delta_0.2_horizontal} intensity and time scales between.

\section{\label{sec:Analysis}Analysis and Conclusions}

We extended the Davies et al. model by incorporating transport networks to consider population displacement over long distances in the city, and applied this extended model to a typical Latin American city configuration. To choose the targets and regions more prone to be attacked, we based our model on the first days of rioting activity from Santiago, Chile, during October 2019. Most of the disorders and criminal activity were highly concentrated around subway stations, and the most active spot was the best--connected region in the city.

Our model’s simulation results were quite surprising because the assumed number of police forces was directly dependent on the total number of rioters. This behavior leads to a counterintuitive outcome. When there is a working transport network, it facilitates the formation of disorder areas, reaching very swiftly a high number of rioters located at precise spots on a relatively small region in the city. On the other hand, if the transport network is not working, the rioting activity takes more time to develop a critical mass, and is extended over a larger area of the city. Contrary to what intuition could indicate, however, it is this second case where the produced disorder has a more considerable intensity, lasting longer and requiring a much larger deployment of police forces. We found that the necessary number is approximately two times larger than for the same situation, with working transport networks.

This counterintuitive behavior is a consequence of the police requirement, which appears in eq.(\ref{eq:requirementorig}). This order forces behavior means that a fast rioter concentration over a small region in space stimulates an equally quick increment on police forces on those regions mentioned above. Once those initial disorders are controlled, and because the total number of police forces cannot decrease, those forces will start to quickly move around the city, as they are required. This strength of resources allows the police forces to effectively regain control in any other area where disorder is formed. On the other hand, when there is no working transport network, this accumulation of forces is slower and more spatially extended, meaning that they take more time to regain control.

An interesting characteristic of the spatial propagation of rioting activity is that it moves, forming well-defined intensity fronts of traveling waves. This kind of behavior has previous been predicted by Berestyki \cite{berestycki2015model, bonnasse2018epidemiological, berestycki2020modeling}. When transport networks are present, those waves’ origin is located at the intersection of the transport lines (the most connected spot on the city) and later moves towards the peripheral and more deprived regions, as shown in FIGS. \ref{fig:rioters_pl_1.5_RC_3000_delta_0.2_horizontal} and \ref{fig:rioters_pl_1.5_RC_3000_delta_0.2_vertical}. This predicted behavior was observed during the riots of Santiago during October 18th 2020, when it was commemorated the first year of the riots of October 2019 \cite{Emol2020-1,Emol2020-2}.

At this point it is important to recall that a fast and efficient disorder extinction is based on two basic premises:

\begin{itemize}
    \item[i] Police forces can increment their numbers as the situation requires, with no limit and as fast as needed.
    
    \item[ii] Those forces can move freely all over the city. The displacement times do not depend on the distances involved.
\end{itemize}

When those two conditions are not fulfilled, transport networks’ presence only increases the speed at which rioters accumulate at the target points, conducing to a much larger and extended disorder activity. Therefore, without this increased accessibility, most of the population is confined to choose to attack only between those targets located nearby they reside in.

There are no quantitative studies on the effects of geographical segregation over crowd behavior in Latin American cities, as far as we are aware. Nevertheless, we can note the work of Lim et al. \cite{lim2007global} and Rutherford et al. \cite{rutherford2014good}.  On those studies, it is shown that boundaries between large-scale segregated areas are historically sources for localized spots of high social tension and civil violence. In addition to the geographical components, there are also social factors; for example, it was found by Bonnasse--Gahot et al \cite{bonnasse2018epidemiological}, for the French riots of 2005, that the most relevant sociological variable used to predict civil violence was the population of young males aged 16--24 with incomplete high--school education. However, a previous formulation proposed by Pires and Crooks \cite{pires2017modeling}, using an agent-based model and incorporating social network analysis and geographical information, identified that the buildup of tension, resulting in the provably civil disorder, is mainly decreased by improving employment possibilities for the population. However, if only education is increased, the result is a more tense and unstable climate. The previous conclusion could be of particular interest for the case of Chile because there, historically, most riots are concentrated at specific spots of cities and seem to be more the result of organized activity than a spontaneous reaction to deprived conditions \cite{govea1981riot}. If Pires and Crooks' model is correct, these environmental conditions look like an explanation for the civil disorder organization in Chile \cite{somma2017discontent}. This seems to be a compounded effect of a significant increase in the number of higher education students \cite{OECD2012}, and the economy deceleration over the last decade \cite{bcentral2019}, creating a build--up of social tension.

\section{\label{sec:Perspectives}Perspectives and Future Work}

For this work, we used a radically simplified population distribution pattern for our model, which captures only the most basic features of a generic Latin American city. A natural extension for our model would be using real geographical data from a major Latin American city. In this way, the approach demonstrated here could be used to make specific predictions. In particular, it could be of practical utility to anticipate the potential level of riot activity at a given location and the number of police forces needed to control any eventual disorder.

Another model characteristic that could be improved is police behavior. In the present model, police forces can move instantaneously all over the city without any distance penalization. A possibility to overcome this issue could be to use an agent-based model, via which differential travel times could be included, potentially leading to a more realistic outcome from their interaction with the civil population.

Finally, another interesting configuration to study could be government intervention over the communication networks, as occurred during the Arab Spring in 2011 \cite{hassanpour2014media}. In this case, temporal disruption of the internet and telephone lines appeared to amplify the disorder, increasing citizen participation, by fragmenting the global communication network and allowing the formation of numerous smaller and independent groups. In future work, therefore, we could study the effects of information control on the population.

\section*{Acknowledgments} 

C.C. wishes to acknowledge the support of FONDECYT (CL), No. 1200357 and Universidad de los Andes (CL) through FAI initiatives.

\bibliographystyle{ieeetr}

\end{document}